\documentclass[sigconf]{acmart}
\makeatletter
\@ACM@balancefalse
\makeatother
\AtBeginDocument{%
  }


\copyrightyear{2026}
\acmYear{2026}
\setcopyright{cc}
\setcctype{by}
\acmConference[UIST '26]{The 39th Annual ACM Symposium on User Interface Software and Technology}{November 02--05, 2026}{Detroit, MI, USA}
\acmBooktitle{The 39th Annual ACM Symposium on User Interface Software and Technology (UIST '26), November 02--05, 2026, Detroit, MI, USA}
\acmDOI{10.1145/3830398.3830493}
\acmISBN{979-8-4007-2856-3/2026/11}

\begin{document}

\title{Present Drawing: Exploring Avatar-Mediated Drawing Events with Physical Traces}

\author{Kojiro Tanaka}
\orcid{0000-0003-3029-1614}
\email{tanaka_kojiro@cyberagent.co.jp}
\affiliation{%
  \institution{CyberAgent}
  \city{Shibuya}
  \state{Tokyo}
  \country{Japan}}
\affiliation{%
  \institution{The University of Osaka}
  \city{Toyonaka}
  \state{Osaka}
  \country{Japan}}

\author{Takuya Iwamoto}
\orcid{0000-0001-9752-1782}
\email{iwamoto_takuya_xa@cyberagent.co.jp}
\affiliation{%
  \institution{CyberAgent}
  \city{Shibuya}
  \state{Tokyo}
  \country{Japan}}
\affiliation{%
  \institution{The University of Osaka}
  \city{Toyonaka}
  \state{Osaka}
  \country{Japan}}

\author{Yuki Okafuji}
\orcid{0000-0002-9547-3681}
\email{okafuji_yuki_xd@cyberagent.co.jp}
\affiliation{%
  \institution{CyberAgent}
  \city{Shibuya}
  \state{Tokyo}
  \country{Japan}}
\affiliation{%
  \institution{The University of Osaka}
  \city{Toyonaka}
  \state{Osaka}
  \country{Japan}}

\renewcommand{\shortauthors}{Tanaka et al.}

\begin{abstract}
We introduce Present Drawing, an approach for staging an avatar-mediated drawing event that leaves physical traces in the real world. Present Drawing aims to reconfigure conventional drawing events by relaxing their physical constraints. We describe an organizer-oriented design space for such an event along four elements: \textit{Time}, \textit{Site}, \textit{Control Source}, and \textit{Avatar Embodiment}. To ground this concept, we developed a proof-of-concept prototype in which an observer wears a head-mounted display to view an avatar. The avatar's motion is synchronized with a customized pen plotter whose mechanical parts are hidden from the observer. We also developed five application examples spanning live, playback, multi-site, mascot, and system-controlled configurations. We conducted an initial evaluation of one representative playback autograph configuration, showing the basic plausibility of the core Present Drawing experience as an avatar-mediated drawing event and highlighting factors associated with perceiving the avatar as the drawer.
\end{abstract}


\begin{teaserfigure}
  \vspace{-15pt}
  \includegraphics[width=\textwidth]{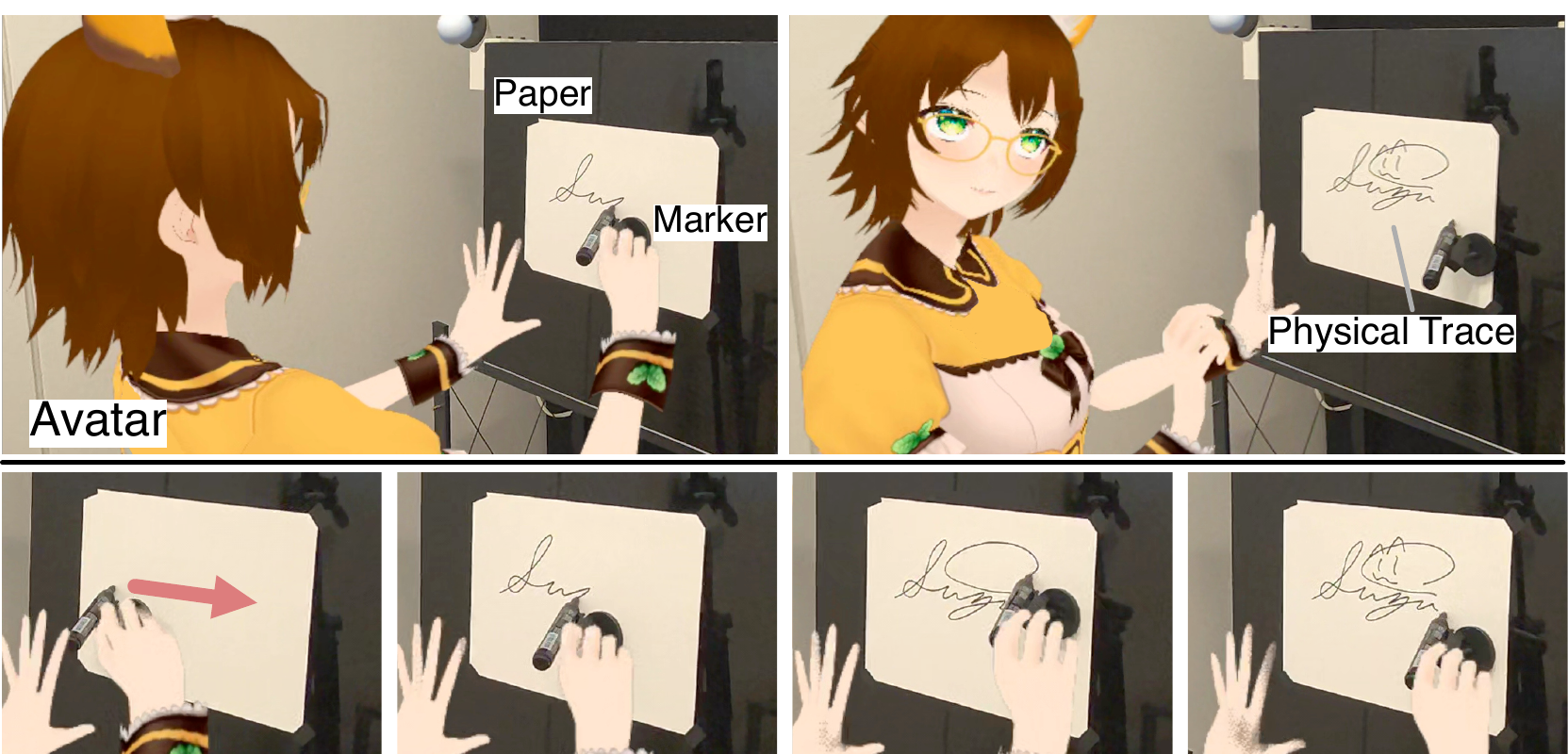}
  \vspace{-15pt}
  \caption{Present Drawing explores an avatar-mediated drawing event with physical traces. For an HMD user, an avatar appears to draw on real paper while the marker moves, producing the corresponding autograph trace.}
  \Description{Present Drawing overview. An avatar stands beside a vertically mounted sheet and appears to draw on real paper. A physical marker moves across the sheet and an autograph gradually appears. The lower sequence shows the marker approaching a blank sheet, drawing the autograph, and leaving a completed trace.}
  \label{fig:teaser}
\end{teaserfigure}


\maketitle

\section{INTRODUCTION}
Drawing is a basic means of expression, communication, and thinking. In some cases, drawings also become valuable artifacts, such as autographs, illustrations, or artworks. Prior research in psychology suggests that the value of an artifact can depend not only on appearance but also on who made it and how it was made \cite{newmanArtAuthenticity2012,gelmanArtifactsEssentialism2013}. This helps explain longstanding and still-popular practices such as talent autograph sessions and live illustration performances, in which an observer watches a specific person produce a drawing in real time; in this paper, we refer to such practices as drawing events.

Traditionally, drawing events have involved physical constraints, such as requiring the drawer and observer to be together in person. At the same time, mixed reality (MR) and avatar research have explored ways to relax these constraints in real-world interaction by placing remote collaborators, represented as avatars, in local spaces \cite{piumsomboonMiniMe2018,jonesBelongingThere2021,yuAvatarsTeleconsultation2021} and, more recently, by allowing virtual avatars to affect physical objects \cite{iharaHoloBots2023,leePhysicalVirtualTable2018,fabreKuroko2024,leeWobblyTable2016}. These works suggest the possibility of coupling an avatar's motion with physical marks that emerge in the real world, opening new possibilities for drawing events beyond their conventional physical form. However, few avatar-mediated approaches have treated drawing events themselves as a primary target domain.

In this paper, we propose Present Drawing, an approach for staging an avatar-mediated drawing event that leaves physical traces in the real world. Present Drawing aims to reconfigure drawing events themselves by relaxing their physical constraints. We present a design space for Present Drawing along four elements: \textit{Time}, \textit{Site}, \textit{Control Source}, and \textit{Avatar Embodiment}. These design-space elements are framed from the perspective of an event organizer who wants to stage a drawing event. Through them, the organizer can consider whether a drawing event is staged live or presented in playback form, whether it is staged at a single site or across multiple sites, whether the drawing is driven by a human operator or by system-controlled motion, and what kind of body is perceived as the drawer.

To ground this concept, we developed a proof-of-concept prototype for Present Drawing. In the prototype, an observer wears a head-mounted display (HMD) to view an avatar whose motion is synchronized with a customized Computer Numerical Control (CNC) pen plotter. The customized plotter hides its actuation mechanism and leaves mainly the marker visible, allowing the marker to generate physical strokes while remaining visually coupled with the avatar's hand. As shown in Figure~\ref{fig:teaser}, this setup makes the avatar appear to draw directly on paper in the real world with a physical marker. Using this prototype, we developed five application examples based on the design space: Live Autograph Session, Moment Message Capsule, Multi-Site Illustration Performance, Mascot Autograph Greeting, and System-Controlled Character Autograph. In addition, we conducted an initial evaluation on one representative configuration of Present Drawing to examine whether its core experience could be perceived as an avatar-mediated drawing event and to highlight factors that help an observer feel that the avatar itself is doing the drawing.

The primary contribution of this paper is the concept of Present Drawing for staging an avatar-mediated drawing event, together with an organizer-oriented design space. We support this contribution with a proof-of-concept prototype, five application examples, and an initial evaluation. Concretely, the contributions are as follows:
\vspace{-6pt}
\begin{enumerate}
  \item A concept of Present Drawing to stage an avatar-mediated drawing event with physical traces, and a design space for expanding conventional drawing events through avatars from an organizer's perspective.
  \item A proof-of-concept prototype featuring a hidden magnetically actuated pen plotter that generates physical strokes while maintaining visual coupling between the avatar's hand and the visible marker, and five application examples that make the organizer-oriented design space concrete.
  \item An initial evaluation of one representative configuration, examining whether its experience can be perceived as an avatar-mediated drawing event and highlighting factors that support perceiving an avatar as a drawer.
\end{enumerate}
\vspace{-10pt}

\section{RELATED WORK}
\subsection{Physical Trace Generation Systems}

Among the technical foundations for extending drawing beyond co-located settings, one important line concerns how traces are physically generated. Research on physical trace generation has taken several forms. One approach uses robot manipulators. Within this line, some studies focus on artistic rendering, painting, calligraphy, and related painterly mark-making systems \cite{tressetPortraitDrawingPaul2013,scaleraWatercolourRoboticPainting2018,sunCalligraphyRobotCallibot2013,deussenFeedbackGuidedStroke2012,schaldenbrandFRIDACollaborativeRobot2022}, while others explore more interactive or adaptive robot-mediated drawing through eye-tracked interaction, impedance-controlled arbitrary-surface drawing, and closed-loop 3-D drawing \cite{scaleraHumanRobotInteractionEye2021,songArtisticPenDrawingArbitrary2018,liuRobustRobotic3DDrawing2022}.

Another approach uses CNC-based mechanisms. Many such systems move a drawing tool mounted on the CNC head across the working surface, for example in direct CNC drawing, low-cost plotter implementations, and CNC-based material drawing \cite{liDirectImmediateDrawing2017,younsDesignImplementationCNC2024,liuCNCAnodizedTitanium2023}. Within this pen-based line, some systems explore alternative ways of driving the pen, for example by using magnetic coupling to move a pen on the top surface \cite{yamaokaDePENdAugmentedHandwriting2013} or by adding pen-pressure control to a pen plotter \cite{kurokiCalligraphyFabricatablePen2022}.

Among the physical trace generation systems introduced above, some studies explore remote handwriting communication, in which handwritten messages are physically delivered to a distant recipient \cite{sonTeleHandwriter2017,kimEffectHandwriting2019}. These works are important here because they show that a physically persistent handwritten trace can function as a communication medium across distance. In this sense, they suggest how physical trace generation can extend writing beyond co-located settings.

However, these works primarily treat handwritten traces as messages to be physically delivered, rather than treating the production of those traces itself as an event to be witnessed. By contrast, we explore how drawing can be staged as an event in which a drawer is perceived to produce a physical trace while the act of production and the emerging trace are experienced together.

\subsection{Drawing and Embodiment in AR/MR}

Augmented reality (AR) and MR have often been used to relax physical constraints such as co-presence, space, and embodiment. One relevant line of work has explored drawing as an interaction resource. For example, AR sketching systems support in-situ ideation and prototyping \cite{aroraSymbiosisSketch2018,gasquesPintAR2019,kwanMobi3DSketch2019}, while remote collaboration systems use strokes, annotations, and related visual cues to indicate intent, guide attention, and support coordination across distance. Such cues are overlaid on a shared view of the scene \cite{gauglitzTouchRemoteWorld2014,fakourfarStabilizedAnnotations2016,kimComparingPointingDrawing2013,teoHandGesturesVisualAnnotation2018} or shared on physical surfaces through projection \cite{junuzovicIllumiShare2012}.

Another relevant line of work has explored the use of avatars as a communication medium for real-world interaction across distance. Examples in this broader space include teleconsultation avatars with varied embodiment techniques \cite{yuAvatarsTeleconsultation2021}, avatar-guided digital representations of physical task objects \cite{tianVirtualReplicas2023}, avatar-overlaid telepresence robots \cite{jonesBelongingThere2021}, and adaptive miniature avatars \cite{piumsomboonMiniMe2018}. 

At the intersection of these two lines are systems that combine avatars and drawing within the same real collaborative setting. Blended Whiteboard supports avatar-mediated remote collaboration in which users discuss ideas around local physical whiteboards while viewing and interacting with virtual sketches overlaid in MR \cite{gronbaekBlendedWhiteboard2024}. MRMAC similarly supports asymmetric MR collaboration in which remote users are represented as avatars and communicate through visual cues including free-form drawing annotations \cite{zamanMRMAC2023}. SurfShare shares the physical surface itself between remote head-and-hand avatars, demonstrating applications such as sketching on the shared surface in MR \cite{huangSurfShare2023}. Taken together, these studies suggest the possibility of avatar-mediated drawing interaction that can extend beyond conventional physical constraints.

These systems overlap with our goal by jointly presenting remote collaborators and drawing surfaces. However, they primarily use avatars and drawings to support discussion and collaboration. We position the presented avatar body itself as the drawer and focus on events in which an observer both witnesses the drawing process and receives a persistent physical artifact produced through the event. Prior work on tangible data souvenirs suggests that physical artifacts can serve as mementos of an experience \cite{petrelliTangibleDataSouvenirs2017}; such artifacts can be kept, gifted, displayed, or used as merchandise after the event.

\vspace{-3pt}

\subsection{MR Systems with Physical Effects}

A separate line of MR research examines how virtual content can be coupled to physical change in the environment. Examples include systems for manipulating actuated objects \cite{kasaharaExTouch2013}, linking virtual sketches with physical behavior \cite{suzukiRealitySketch2020,kaimotoSketchedReality2022}, and secretly propagating virtual manipulations into the real world through hidden robots \cite{kariRealityPromises2025}. This line also includes systems in which a virtual avatar appears to affect physical objects in the real world.

The Wobbly Table used subtle movement of a shared real-virtual table to increase social presence with a virtual avatar \cite{leeWobblyTable2016}, while the Physical-Virtual Table showed that a virtual avatar's apparent influence on physical objects affects social interaction \cite{leePhysicalVirtualTable2018}. Together, these studies suggest the possibility that visible virtual avatars can change how physical effects are interpreted in social settings.

The Kuroko Paradigm further examines perceived causality under hidden actuation by overlaying a virtual avatar on a robot arm so that resulting physical effects are perceived as caused by the avatar rather than by a visible machine \cite{fabreKuroko2024}. This provides a useful reference because minimizing visible mechanical cues can likewise shape whether a physical effect is perceived as caused by the avatar.

HoloBots allows remote collaborators, represented holographically and embodied through mobile robots, to affect local objects and can even support attachment-based drawing scenarios \cite{iharaHoloBots2023}. This makes it relevant as an example of how avatar-like remote embodiment can extend to physical action and even drawing-related scenarios in the local environment.

Taken together, these works show that virtual avatars can be perceived as agents of physical effects in the real world. However, they stop short of explicitly framing these capabilities as a drawing event that can be staged beyond conventional physical constraints. We explore how avatar-linked physical effects can be used to stage an avatar-mediated drawing event beyond conventional physical constraints while producing physical traces in the real world.

\begin{figure*}[h]
  \centering
  \includegraphics[width=0.92\textwidth]{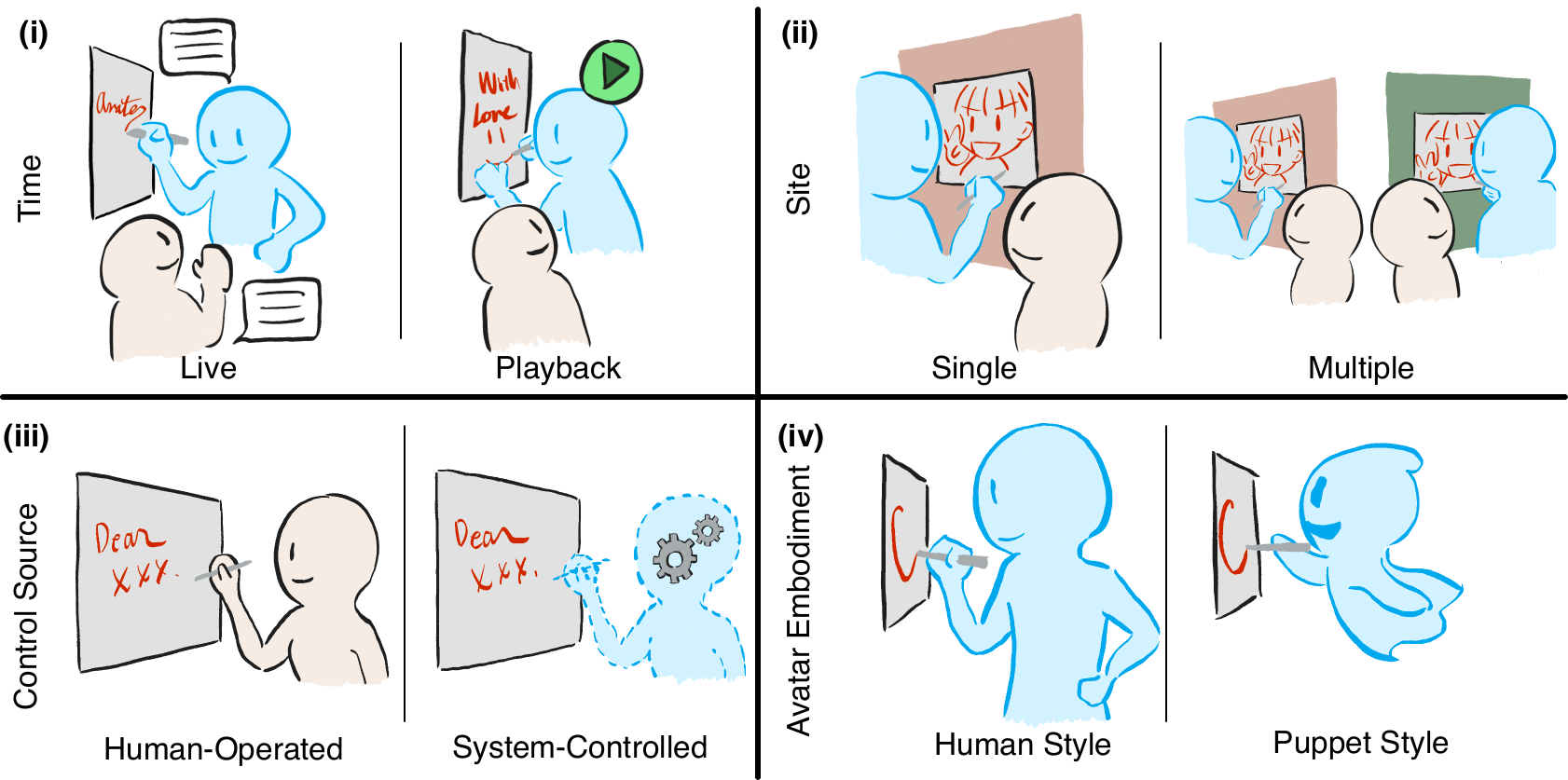}
  \vspace{-10pt}
  \caption{Organizer-Oriented Design Space. Present Drawing varies along four axes: \textit{Time} (Live/Playback), \textit{Site} (Single/Multiple Sites), \textit{Control Source} (Human-Operated/System-Controlled), and \textit{Avatar Embodiment} (e.g., Human Style/Puppet Style). \textit{Time}, \textit{Site}, and \textit{Avatar Embodiment} describe event presentation, whereas \textit{Control Source} describes what drives the drawing action.}
  \Description{This figure presents the four axes of the design space as four sub-figures, (i) to (iv). Sub-figure (i) shows the Time axis, with two sketches comparing live and playback drawing events. In the live case, an observer watches an avatar draw in real time and communicate; in the playback case, a recorded drawing is replayed later for an observer. Sub-figure (ii) shows the Site axis, with two sketches comparing single-site and multi-site presentation. One shows a drawing event at one location, and the other shows the same event reproduced at multiple locations. Sub-figure (iii) shows the Control Source axis, with two sketches comparing human-operated and system-controlled drawing. One shows drawing driven by a human performer, and the other shows drawing generated by the system. Sub-figure (iv) shows the Avatar Embodiment axis, with two sketches comparing different presented bodies for the drawer. One avatar is human-like, and the other is a stylized puppet-like character.}
  \label{fig:designspace}
\end{figure*}

\section{PRESENT DRAWING}
\label{sec:present-drawing}
\subsection{Concept}

Present Drawing is an approach for staging drawing events in which an avatar is perceived as a drawer while physically persistent traces are produced in the real world. It couples the avatar's drawing motion, the motion of a physical drawing tool, and the appearance of material traces on a physical surface. Through this coupling, an observer experiences the action and the resulting trace as a single event. In this paper, we focus on a local setup in which an avatar appears to draw with a marker on paper in the real world while a co-located observer watches the physical trace emerge.

\subsection{Design Space}
Present Drawing aims to extend traditional physical drawing events, such as fan autograph sessions and live illustration performances, by replacing the visible drawer with an avatar. For the design space of Present Drawing, we focus on how such events can be staged more flexibly despite physical constraints, rather than on the individual interactions of the drawer and observer. We therefore describe its design space from the perspective of an event organizer and derive its axes from prior work. Computer-supported cooperative work (CSCW) frameworks have long used time and place to characterize mediated collaborative activity \cite{johansenGroupware1988,leeMatrixModel2015}; applied to physical drawing events, this motivates \textit{Time} (whether the event is live or playback) and \textit{Site} (whether it is staged at a single site or multiple sites). Prior work on VTubing describes choices in avatar representation and control \cite{kimVTubersAtelier2025}, while AR remote collaboration research shows that avatar appearance affects social presence and user perception \cite{yoonEffectAvatarAppearance2019}; applied here, these motivate \textit{Control Source} (whether the drawing is human-operated or system-controlled) and \textit{Avatar Embodiment} (what kind of body is presented as the drawer). Taken together, these perspectives yield the four-axis design space shown in Figure~\ref{fig:designspace}.

\vspace{-2pt}
\subsubsection{\textit{Time}}
\textit{Time} describes whether a drawing event is presented live or in playback as shown in Figure \ref{fig:designspace}-(i). In the live mode, the drawing unfolds in real time and the drawer and observer can interact, as in a conventional in-person event. In the playback mode, recorded drawing actions are presented later so that a past drawing can be experienced again as an event rather than only as a finished drawing. This allows the moment of creation to be replayed together with the resulting trace.

\vspace{-2pt}
\subsubsection{\textit{Site}}
\textit{Site} describes whether a drawing event is staged at a single site or across multiple sites as shown in Figure \ref{fig:designspace}-(ii). Conventional physical drawing is tied to one location because the drawer's body can be in only one place at a time. With Present Drawing, the same drawing event can instead be delivered in parallel to multiple distant locations while leaving physical traces at each site.

\vspace{-2pt}
\subsubsection{\textit{Control Source}}
\textit{Control Source} refers to the source from which the avatar's motion and the associated drawing path are derived. As shown in Figure \ref{fig:designspace}-(iii), it has two modes: human-operated and system-controlled. In the human-operated mode, a human operator provides the body motion and drawing path, which are then used to present the drawing action and generate the physical trace at the observer's site. In the system-controlled mode, the system determines both the presented drawing action and the drawing path that generates the physical trace, allowing developers to create and edit drawing events without requiring a human operator to perform them each time.

\vspace{-2pt}
\subsubsection{\textit{Avatar Embodiment}}
\textit{Avatar Embodiment} describes what kind of body is presented as the drawer. Unlike conventional drawing events, in which the body visible as the drawer is ordinarily the same body that physically produces the drawing, Present Drawing allows the organizer to choose the presented body independently of \textit{Control Source}. The drawer can therefore range from a figure close to the real performer to a stylized character, and from human scale to puppet-sized embodiments, while still leaving physically persistent traces on paper as shown in Figure \ref{fig:designspace}-(iv).

\vspace{-4pt}
\section{PROOF-OF-CONCEPT PROTOTYPE}
\label{sec:system-design}
To make the design space concrete, we developed a proof-of-concept prototype of Present Drawing. The prototype consisted of two phases: an authoring phase and a performance phase. During the authoring phase, an event package consisting of synchronized avatar motion and stroke traces was captured. During the performance phase, the event package was presented as the avatar's drawing action while the same stroke traces were materialized as physical marks on paper. To strengthen the impression that the avatar was directly drawing with a physical marker, we focused on approaches that minimize visible mechanical cues, such as the Kuroko Paradigm \cite{fabreKuroko2024}. Conventional CNC-based pen plotters can generate physical strokes but generally leave their mechanical components visible \cite{sonTeleHandwriter2017}. Hidden actuation has also been used to move physical objects \cite{leePhysicalVirtualTable2018} and to produce pen motion through magnetic coupling \cite{yamaokaDePENdAugmentedHandwriting2013}. However, among these existing approaches, few address pen plotters that hide their mechanical parts while also featuring a pen-up/down mechanism. We therefore developed a self-moving pen plotter that conceals both mechanisms, leaving mainly the paper and marker visible. We first describe this hardware and then explain the two phases.

\subsection{Self-Moving Pen Plotter}

\begin{figure}[h]
  \centering
  \includegraphics[width=0.9\linewidth]{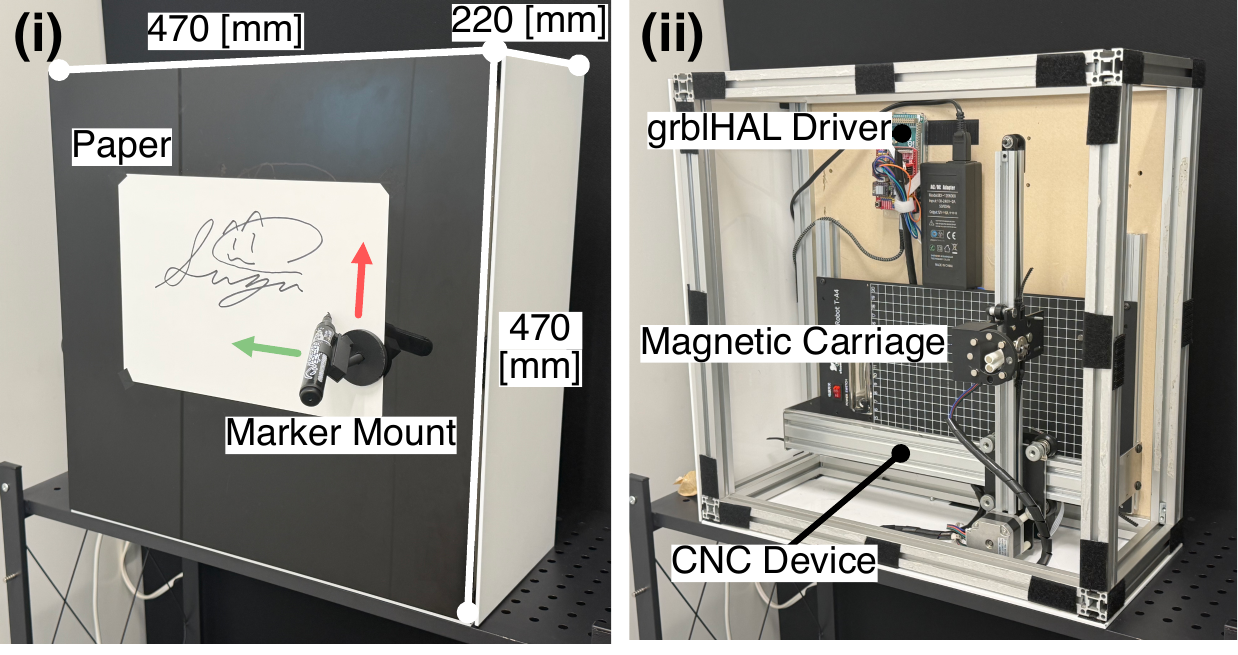}
  \vspace{-10pt}
  \caption{Self-Moving Pen Plotter. (i) Observer-side view, exposing mainly the paper and marker mount. (ii) Internal view with the CNC device and magnetic carriage.}
  \Description{This figure contains two sub-figures, (i) and (ii). Sub-figure (i) is the observer-side view of the pen plotter. The device appears as a box-shaped enclosure with a black front drawing surface, a mounted sheet of paper, and a visible marker mount; from this side, the observer mainly sees the paper and marker rather than the mechanism. Sub-figure (ii) is the internal view of the pen plotter. The enclosure is opened to show the CNC device, rails, magnetic carriage, and controller hardware behind the drawing surface, and these hidden components produce the visible marker motion.}
  \label{cnc}
\end{figure}

The self-moving pen plotter was housed in an aluminum-frame enclosure measuring 470 $\times$ 470 $\times$ 220 [mm], with its sides covered by white acrylic panels, as shown in Figure~\ref{cnc}-(i). Its front drawing surface was a black acrylic plate (thickness: 2 [mm]) on which A4 Kent paper was mounted. A marker mount moved across the paper surface and drew directly on the sheet.

\begin{figure}[h]
  \centering
  \includegraphics[width=0.9\linewidth]{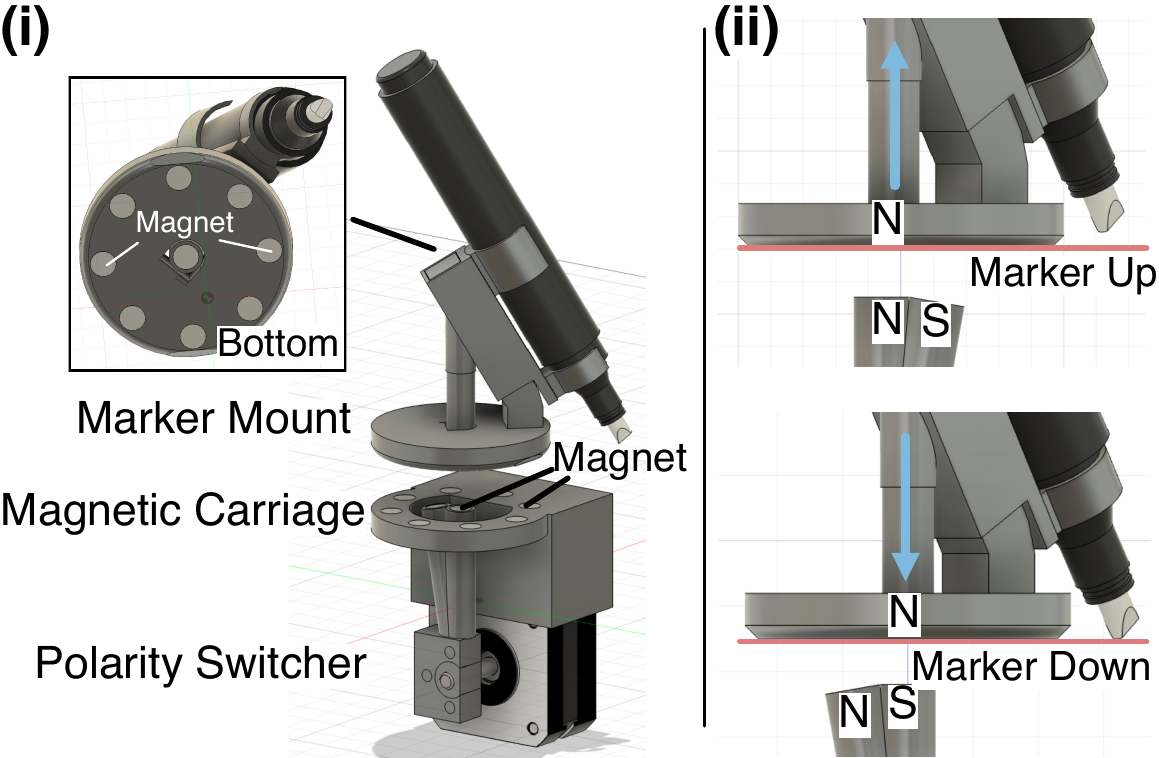}
  \vspace{-10pt}
  \caption{Magnetic Actuation Mechanism. (i) Magnets couple the marker mount to a magnetic carriage. (ii) A polarity switch lifts the marker for travel and lowers it for drawing.}
  \Description{This figure contains two sub-figures, (i) and (ii). Sub-figure (i) is an exploded view of the magnetic actuation mechanism. The visible marker mount sits above the drawing surface, and the magnetic carriage and polarity-switching unit sit below it, so the marker mount is driven through magnetic coupling across the surface. Sub-figure (ii) shows the two states of the magnetic actuation mechanism. In one state, the marker is lifted away from the paper for travel; in the other, it is lowered onto the paper for drawing, and reversing magnetic polarity switches between these states.}
  \label{mount}
  \vspace{-5pt}
\end{figure}

Inside the enclosure, we installed a CNC device (working area: 297 $\times$ 210 [mm]) and a grblHAL driver\footnote{grblHAL Driver: \url{https://github.com/grblHAL}} (Arduino Due), as shown in Figure~\ref{cnc}-(ii). The CNC device used two stepper motors to move the head in a plane. A magnetic carriage attached to the CNC head then transferred the head's motion to the marker mount through the acrylic plate. As shown in Figure~\ref{mount}-(i), magnets on the carriage and on the bottom of the marker mount coupled the two components so that the marker moved with the CNC head while the actuation mechanism remained mostly hidden from the observer.

The marker mount also contained a central magnet, and the magnetic carriage included a magnet-equipped polarity switcher linked to an additional stepper motor mounted on the CNC head. Because the switcher placed the north and south poles side by side, rotating it changed which pole faced the marker-mount magnet. As shown in Figure~\ref{mount}-(ii), like poles repelled each other and lifted the marker, whereas opposite poles attracted and lowered it.

This self-moving pen plotter can operate with the drawing surface oriented either vertically or horizontally. For the prototype presented in this paper, we used the vertical configuration shown in Figure~\ref{cnc}, as vertical drawing is familiar in event formats such as display-panel signing and live illustration on canvas.

\vspace{-4pt}
\subsection{Process}
This subsection introduces the authoring phase and the performance phase. Both phases used the same Windows server PC (Intel Core i9-14900HX CPU, GeForce RTX 4070 Laptop GPU, 32 GB RAM) and an HMD (Meta Quest 3). We developed separate Unity applications for the server PC and the HMD, with the HMD application using video pass-through.

\vspace{-2pt}
\subsubsection{Authoring Phase}
\label{sec:authoring-phase}
This phase produced an event package consisting of synchronized avatar motion and stroke traces. In the human-operated mode,\label{sec:human-operated-mode} a human provided the source motion and stroke data. In the default configuration, avatar motion was captured with the HMD's body-tracking function (Meta Movement SDK\footnote{Meta Movement SDK: \url{https://github.com/oculus-samples/Unity-Movement}}), and stroke traces were captured with our custom drawing application on an approximately A4-sized tablet (iPad Pro, 13-inch, Apple Silicon M4). Because the tablet size closely matched the CNC working area, the application captured strokes in coordinates corresponding to the plotter workspace, allowing the CNC to later reproduce them at nearly the same position and scale.

As shown in Figure~\ref{process}-(i), we matched the authoring setup to the plotter by preparing an aluminum-frame enclosure of the same size and placing the tablet so that its drawing area fell within the marker mount's working area. Two HMD hand controllers were then fixed to the top of the enclosure to serve as spatial anchors for the HMD, allowing avatar motion to be captured in the enclosure's coordinate frame.

While wearing the HMD in video pass-through mode, the drawer drew on the tablet, capturing avatar motion and stroke data together. Because the captured stroke positions corresponded to the positions at which the CNC would later reproduce them, the captured hand positions were also aligned with the marker-mount positions. The motion and stroke data were then sent from the HMD and tablet to the server PC via User Datagram Protocol (UDP) and stored there with synchronized timestamps as an event package. The application examples in Sections~\ref{sec:app-live-signing}, \ref{sec:app-moment-message-capsule}, and \ref{sec:app-multi-site-illustration} used this mode in the default human-scale form, while the application example in Section~\ref{sec:app-mascot-autograph-greeting} used a puppet-like variant authored through hand tracking rather than full-body capture.

In the system-controlled mode,\label{sec:system-controlled-mode} the server PC generated both the stroke traces and the corresponding avatar motion instead of the HMD drawer. Pre-authored stroke data for each character were combined on the server PC according to an input string. Avatar motion was then synthesized with an inverse kinematics library (Final IK\footnote{Final IK: \url{https://assetstore.unity.com/packages/tools/animation/final-ik-14290}}) so that the hand and upper body followed the generated stroke path. By temporally aligning the generated stroke data and synthesized motion, the server PC produced an event package with synchronized timestamps. This mode was later used in the application example in Section~\ref{sec:app-agentic-character-autograph}.

\vspace{-2pt}
\subsubsection{Performance Phase}
\begin{figure}[h]
  \centering
  \includegraphics[width=\linewidth]{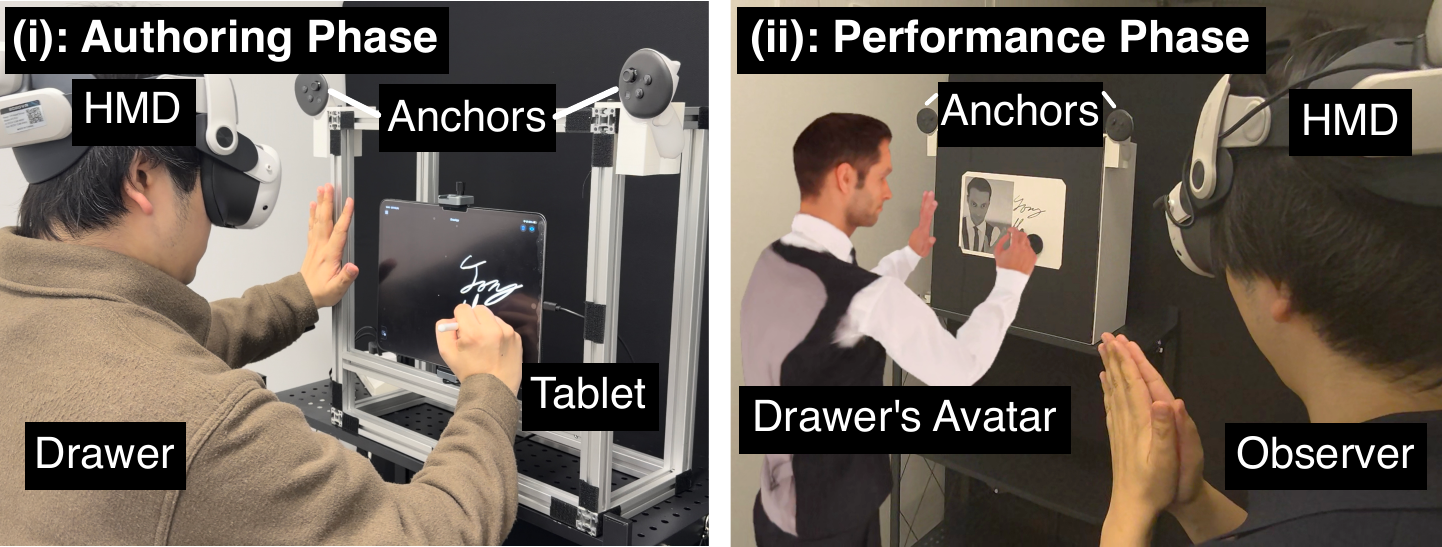}
  \vspace{-10pt}
  \caption{Authoring and Performance Phases. (i) Drawer motion and stroke paths are captured in a plotter-aligned workspace. (ii) The drawing event is staged for the observer: the avatar appears beside the plotter while the corresponding paths are materialized on paper.}
  \Description{This figure contains two sub-figures, (i) and (ii). Sub-figure (i) is the authoring phase. A person wearing an HMD stands in front of a framed workspace and draws on a tablet mounted inside it, and the system captures body motion and stroke paths in a workspace aligned with the plotter. Sub-figure (ii) is the performance phase. An observer wearing an HMD stands in front of the pen plotter and sees an avatar beside the paper; as the avatar appears to draw, the marker moves on the real sheet and reproduces the path physically.}
  \label{process}
  \vspace{-5pt}
\end{figure}

The performance phase showed a drawing event to an observer wearing the HMD with the self-moving pen plotter, as shown in Figure~\ref{process}-(ii). To align the observer-side coordinate frame with the authoring frame, the observer's HMD hand controller was mounted on the self-moving pen plotter as a shared spatial anchor. Avatar motion data were then sent from the server PC to the HMD via UDP, and the avatar was rendered in the real-world view relative to this shared anchor. For stroke motion, the server PC sent marker-path commands to the plotter through a grblHAL serial link to drive the marker mount along the corresponding path. 

The server PC switched the avatar motion timeline between the original motion-data timestamps and marker-position-based estimated timestamps to account for the CNC device's limited dynamic response at high acceleration. When the plotter was idle, the avatar followed the original motion-data timestamps. During drawing, the server PC read the marker position through the grblHAL interface and estimated the corresponding timestamp along the commanded stroke path. The avatar was then driven with this estimated timestamp so that the avatar motion stayed aligned with the physical marker motion in both the live and playback modes described below. Through this process, the drawing-event package was reproduced in the performance phase.

The same pipeline supported both the live and playback modes of \textit{Time}. In the live mode, motion and stroke data produced in the authoring phase were streamed directly to the performance phase in real time. In the playback mode, a previously stored event package was later read from the server PC and presented to the observer. In the live mode, however, using the estimated timestamps during drawing could temporarily shift the avatar motion away from the performer's latest real-time pose. After each drawing segment, the server PC therefore switched the avatar motion timeline back to the original motion-data timestamps and updated the avatar to the latest live pose before post-drawing communication continued.

\vspace{-4pt}
\section{\texorpdfstring{APPLICATION EXAMPLES}{APPLICATION EXAMPLES}}
\label{sec:applications}
\begin{figure*}[t]
  \centering
  \includegraphics[width=0.96\linewidth]{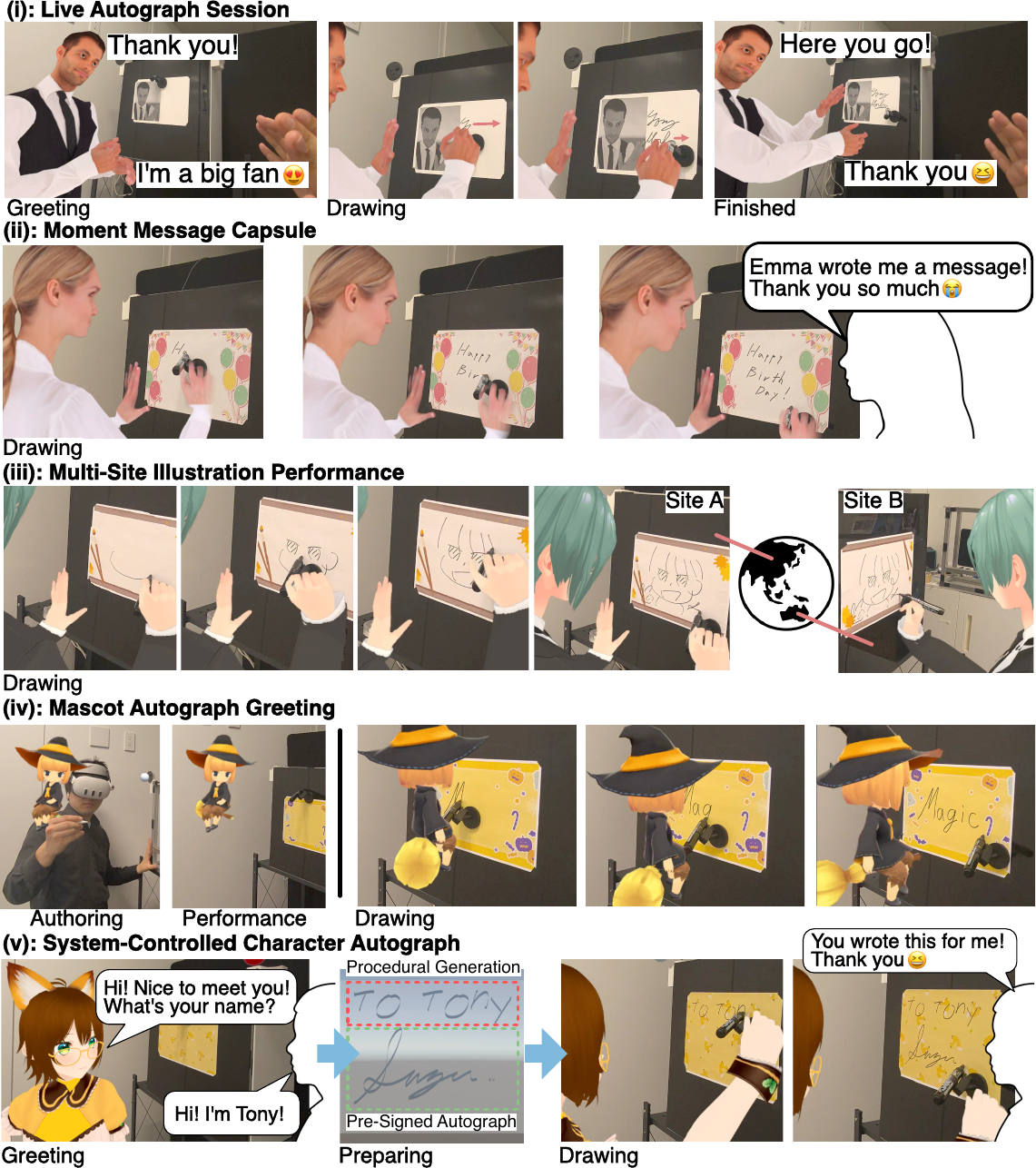}
  \vspace{-10pt}
  \caption{Application Examples Across Design Space. (i) Live Autograph Session: An avatar interacts with an attendee while producing a physical autograph. (ii) Moment Message Capsule: A prerecorded message is replayed later as a witnessed drawing event. (iii) Multi-Site Illustration Performance: The same performance is reproduced at multiple venues. (iv) Mascot Autograph Greeting: A palm-sized mascot appears to draw on its own. (v) System-Controlled Character Autograph: A personalized autograph is generated and drawn without a live human operator.}
  \Description{This figure contains five sub-figures, (i) to (v), one for each application example. Sub-figure (i) is a live autograph session, where a life-sized avatar greets an attendee, writes on a mounted portrait sheet, and presents the finished autograph, showing a real-time interactive signing experience. Sub-figure (ii) is a moment message capsule, where a prerecorded birthday message is written gradually on a decorated sheet and later received as a witnessed drawing event, so the message appears over time rather than only as a finished result. Sub-figure (iii) is a multi-site illustration performance, where the same drawing action is reproduced at different venues so that similar illustrations appear at Site A and Site B, showing one performance staged across multiple locations. Sub-figure (iv) is a mascot autograph greeting, where a small witch-like character is puppeteered during authoring and later appears beside the board during performance, seeming to move the marker and draw by itself. Sub-figure (v) is a system-controlled character autograph, where a conversational character receives a participant's name, combines it with a pre-authored signature, and draws the resulting personalized autograph, generated without a live human operator.}
  \label{App}
\end{figure*}

Building on the design space in Section~\ref{sec:present-drawing}, we developed five application examples of Present Drawing. The application examples in Sections~\ref{sec:app-live-signing} and \ref{sec:app-moment-message-capsule} primarily instantiate \textit{Time}, while the application examples in Sections~\ref{sec:app-multi-site-illustration}, \ref{sec:app-mascot-autograph-greeting}, and \ref{sec:app-agentic-character-autograph} instantiate \textit{Site}, \textit{Avatar Embodiment}, and \textit{Control Source}, respectively. Sections~\ref{sec:app-live-signing}--\ref{sec:app-mascot-autograph-greeting} use the human-operated mode, whereas Section~\ref{sec:app-agentic-character-autograph} uses the system-controlled mode.
\newcounter{renderpeopleavatarfn}
\newcounter{vroidavatarfn}

\vspace{-4pt}
\subsection{Live Autograph Session}
\label{sec:app-live-signing}
This application example instantiates the live form of \textit{Time}: an avatar and an attendee interact in real time while the avatar produces a physical autograph. A familiar example is a talent autograph session. As shown in Figure~\ref{App}-(i)\footnote{\setcounter{renderpeopleavatarfn}{\value{footnote}}The avatar was sourced from RENDERPEOPLE Free 3D People: \url{https://renderpeople.com/free-3d-people/}.}, the attendee can communicate with the talent while receiving a physically persistent autograph. Because the drawing act is embodied by an avatar, organizers can stage this format beyond an ordinary in-person event, for example by having a talent draw for an attendee at another site while still appearing locally to that attendee. More recently, natively digital talents such as VTubers have also emerged \cite{luMoreKawaii2021,kimVTubersAtelier2025}. Present Drawing suggests that such digital talents can also hold live autograph sessions in the physical world.

\vspace{-4pt}
\subsection{Moment Message Capsule}
\label{sec:app-moment-message-capsule}
This application example instantiates the playback form of \textit{Time} by packaging not only the message itself but also the moment in which it is drawn. Figure~\ref{App}-(ii)\footnotemark[\value{renderpeopleavatarfn}] shows a birthday message being drawn for the recipient. The sender records the drawing motion and stroke data in advance, and the recipient later experiences the recorded drawing as an event rather than receiving only the finished message.

\vspace{-4pt}
\subsection{Multi-Site Illustration Performance}
\label{sec:app-multi-site-illustration}
Live illustration performance is a familiar drawing event because audiences can watch a manga artist or illustrator draw in real time. This application example extends that format across sites: as shown in Figure~\ref{App}-(iii)\footnote{\setcounter{vroidavatarfn}{\value{footnote}}The avatar was adapted from a sample avatar obtained through VRoid Studio: \url{https://vroid.com/en/studio}.}, the same illustration performance can be presented simultaneously at multiple venues while generating a physical drawing at each one. Because the resulting illustration remains as a physical artifact, it could also be offered at each venue as merchandise.

\vspace{-4pt}
\subsection{Mascot Autograph Greeting}
\label{sec:app-mascot-autograph-greeting}
This application example explores a non-human-scale embodiment in which a palm-sized mascot character signs or draws for a visitor. It is inspired by greeting experiences such as those in theme parks, where visitors receive autographs or drawings directly from a character. Figure~\ref{App}-(iv)\footnote{The avatar was adapted from the Unity Asset Store package ``Novice Witch - Haon SD Series'': \url{https://assetstore.unity.com/packages/3d/characters/novice-witch-haon-sd-series-180964}. © Unity Technologies Japan/UCL © Haon} shows a palm-sized witch character producing a drawing. In the authoring phase, the character is controlled in a puppet-like manner through hand tracking so that its motion and the resulting autograph strokes are captured together. During the performance phase, the small witch then appears to move the marker and draw on its own.

\vspace{-4pt}
\subsection{System-Controlled Character Autograph}
\label{sec:app-agentic-character-autograph}
MR research has also explored virtual agents situated in real-world settings \cite{holz2011mira}. Following this direction, we developed an application example in which a system-controlled character appears to draw a personalized autograph. This example instantiates the system-controlled side of \textit{Control Source}: both the avatar motion and the handwritten trace are procedurally generated. Figure~\ref{App}-(v)\footnotemark[\value{vroidavatarfn}] shows this application example. In this application example, strokes for the participant's name are generated from pre-authored character data and combined with a pre-signed autograph to form a personalized handwriting trace. The character then draws along that trace, creating a system-controlled drawing event.

\vspace{-4pt}
\section{INITIAL EVALUATION}
\label{sec:userstudy}

The previous section presented five application examples showing how Present Drawing could expand drawing events. However, such extensions are meaningful only if an observer first recognizes Present Drawing itself as a drawing event. We therefore conducted an initial evaluation to examine whether the core experience of the current Present Drawing prototype could be perceived as an avatar-mediated drawing event. This evaluation focused on a single representative playback condition shared by all participants. Its goal was not to compare the design-space axes of Present Drawing, but to test whether one controlled configuration could support that core perception. We also highlighted factors associated with an observer's sense that the avatar itself is drawing, because this sense is important for Present Drawing to be experienced as a drawing event. This evaluation was approved by The University of Osaka (Reference Number: R1-5-22).

\subsection{Method}
\vspace{-2pt}
\subsubsection{Participants}
We recruited 8 participants (5 female, 3 male) from our institution, excluding members of the authors' laboratory. None had prior knowledge of the evaluation; all were native speakers of Japanese with normal or corrected-to-normal vision.

\vspace{-2pt}
\subsubsection{Evaluation Setup}
\begin{figure}[h]
  \centering
  \includegraphics[width=0.95\linewidth]{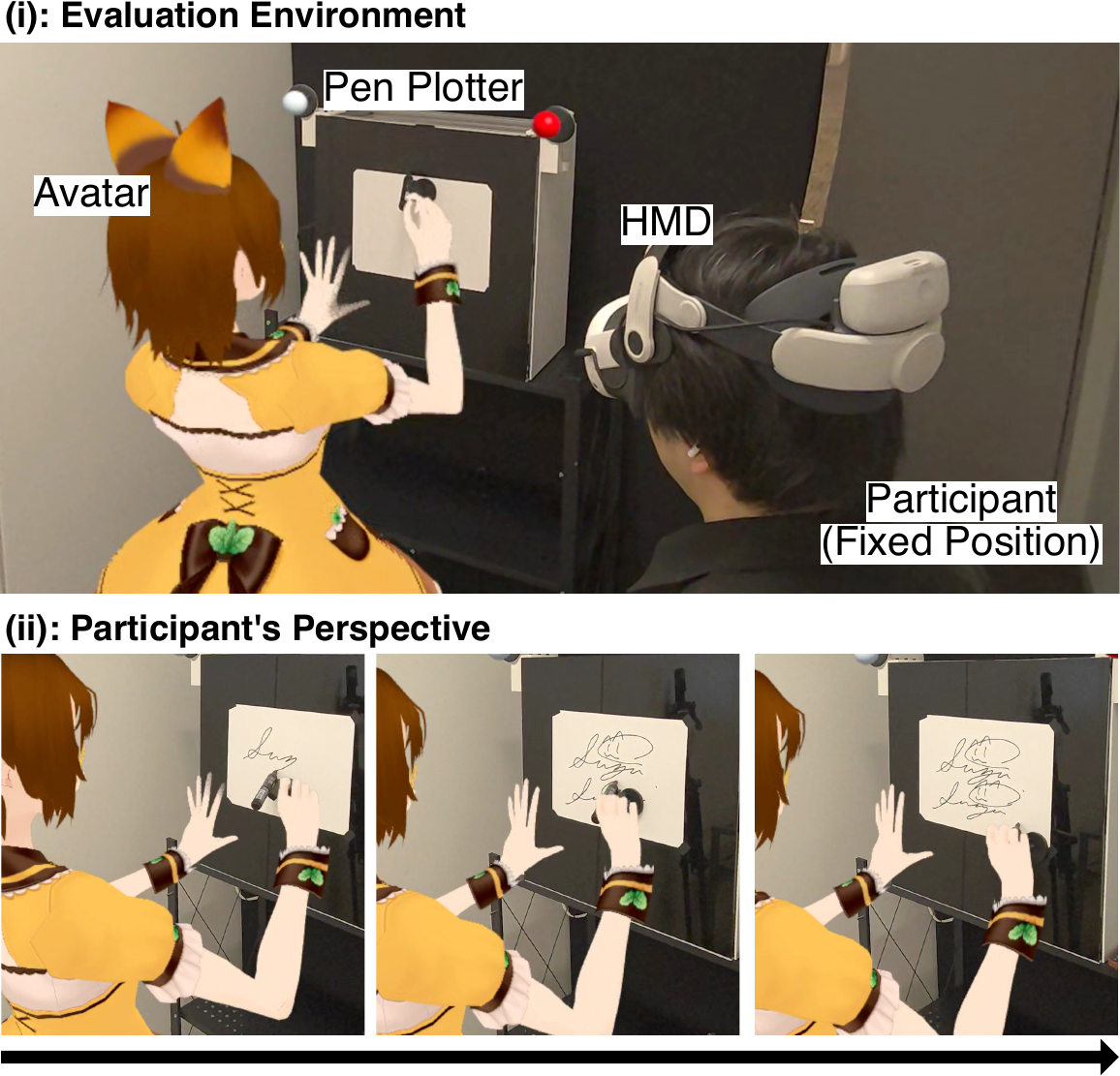}
  \vspace{-10pt}
  \caption{Evaluation Setup and Participant's View. (i) A participant observed the drawing avatar from a fixed position. (ii) Participant's view during the playback autograph session.}
  \Description{This figure contains two sub-figures, (i) and (ii). Sub-figure (i) is the evaluation setup in the laboratory. A participant wearing an HMD stands at a fixed viewing position in front of the pen plotter, and the avatar appears beside the device and faces the drawing surface. Sub-figure (ii) is the participant view during the playback autograph session. The avatar's hand stays near the paper while the physical marker moves across the sheet and the autograph trace gradually appears.}
  \label{setup}
  \vspace{0pt}
\end{figure}

Figure~\ref{setup}-(i) shows the evaluation environment. The evaluation was conducted in a room in our laboratory. We mounted an A4-sized sheet on the self-moving pen plotter and placed it on a shelf so that the center of the sheet was about 1.4~m high. Participants stood at a fixed position about 1.0~m from the plotter and wore an HMD with pass-through enabled to observe the drawing event.

For this evaluation, we used a VTuber autograph event as the drawing-event scenario. The event was a fictitious scenario pre-authored in the authoring phase of Present Drawing and then presented in playback format, with the avatar delivering pre-scripted speech and not responding to user interaction, so that all participants experienced the same event content. In the event, the VTuber greeted the participant and then drew an autograph on the paper twice, as shown in Figure~\ref{setup}-(ii).

\vspace{-2pt}
\subsubsection{Measurements}
We evaluated participants' perceptions of avatar-mediated drawing using several 7-point Likert-scale items. Table~\ref{tab:likert} lists the questionnaire items. Q1--Q3 were custom items designed for Present Drawing to assess whether participants felt that the avatar itself drew the autograph, whether the emerging autograph was experienced as naturally synchronized with the avatar's hand movement, and whether the resulting output felt odd as an autograph. Q4 and Q5 were lightweight presence-related items. Prior work on avatar-mediated physical effects often examines social presence or co-presence \cite{leePhysicalVirtualTable2018,iharaHoloBots2023,fabreKuroko2024,leeWobblyTable2016}; following this line, we included two simple items as a supplementary check of whether the event as a whole conveyed avatar presence in real space and whether that presence changed during drawing. These items were intended to complement the main evaluation of avatar-mediated drawing rather than to provide a full presence assessment. After the questionnaire, we conducted semi-structured interviews to gather qualitative feedback and analyzed these interview comments using thematic analysis \cite{braunUsingThematicAnalysis2006}.

\begin{table}[h]
\centering
\small
\caption{Likert-Scale Questionnaire Items Used in the Evaluation (1 = Strongly Disagree, 7 = Strongly Agree)}
\Description{Questionnaire items used in the evaluation. The table lists five 7-point Likert-scale items about whether the avatar felt like the drawer, whether the emerging lines matched the avatar's hand movement, whether the autograph felt odd, whether the avatar felt present, and whether that sense of presence became stronger while it was drawing.}
\label{tab:likert}
\vspace{-5pt}
\begin{tabular}{p{0.08\linewidth} p{0.82\linewidth}}
\hline
\textbf{Item} & \textbf{7-point Likert-scale item} \\
\hline
Q1 & Whether the autograph drawn in front of me felt as if the avatar itself drew it. \\
Q2 & Whether it felt as if the lines of the autograph were appearing naturally in sync with the avatar's hand movement. \\
Q3 & Whether the resulting output felt odd as an autograph. \\
Q4 & Whether it felt like the avatar was there during the experiment. \\
Q5 & Whether the feeling that the avatar was there became stronger while it was drawing than when it was not drawing. \\
\hline
\end{tabular}
\end{table}

\vspace{-2pt}
\subsubsection{Procedure}
After signing the consent form, each participant was told that they would observe a VTuber autograph session and was shown the avatar used in the evaluation. Participants then received a brief tutorial on the HMD and were helped to put it on. They experienced the VTuber autograph session, completed the questionnaire in Table~\ref{tab:likert}, and then had a semi-structured interview lasting about 15~minutes. The total session lasted about 30~minutes per participant.

\vspace{-4pt}
\subsection{Results}
\begin{figure}[h]
  \centering
  \includegraphics[width=0.97\linewidth]{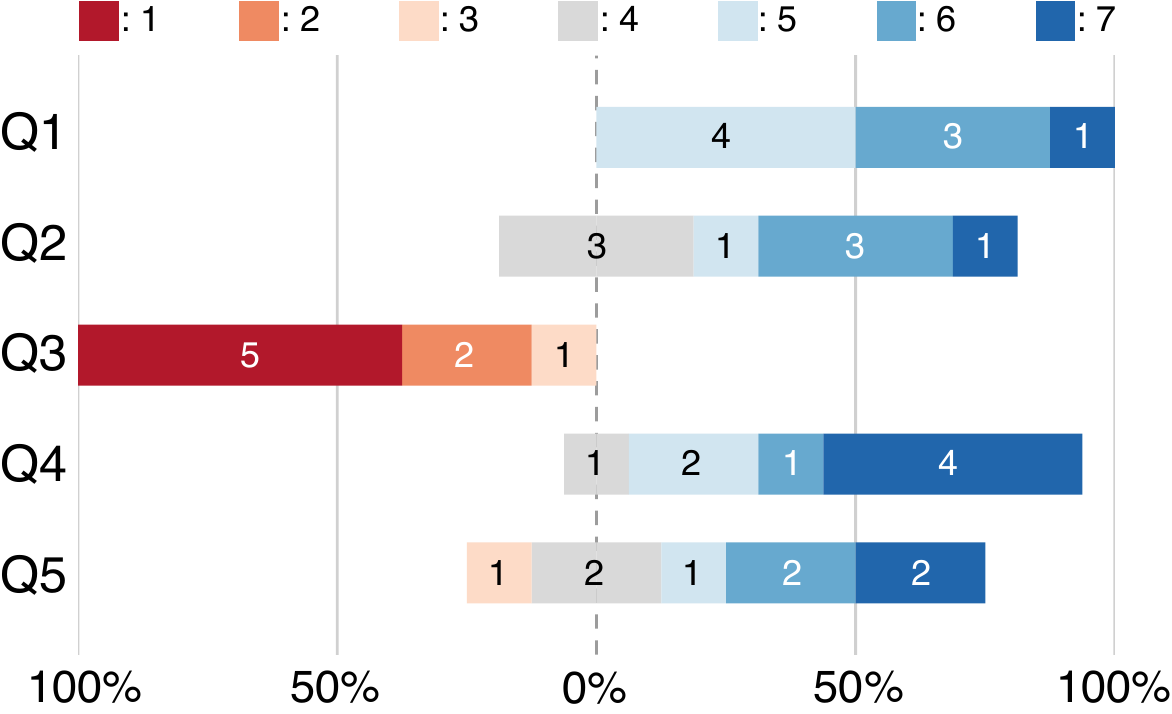}
  \vspace{-10pt}
  \caption{Quantitative Results. Diverging stacked bar charts of the questionnaire responses relative to the midpoint of 4. The vertical axis lists questionnaire items Q1--Q5: Q1 avatar as drawer, Q2 hand--trace synchrony, Q3 autograph oddness, Q4 avatar presence, and Q5 increased presence during drawing. The horizontal axis shows response percentages, with lower ratings extending left and higher ratings extending right from the 0\% point. For visualization, responses at the midpoint of 4 are centered at the 0\% point, with half of the corresponding segment extending to each side. Numbers in the bars indicate participant counts.}
  \Description{The chart shows responses for Q1 through Q5 on a 7-point scale as diverging stacked bars centered at the neutral midpoint of 4. Ratings below 4 extend to the left and ratings above 4 extend to the right. Q1 has four responses at 5, three at 6, and one at 7. Q2 has three responses at 4, one at 5, three at 6, and one at 7. Q3 has five responses at 1, two at 2, and one at 3. Q4 has one response at 4, two at 5, one at 6, and four at 7. Q5 has one response at 3, two at 4, one at 5, two at 6, and two at 7.}
  \label{result}
  \vspace{-5pt}
\end{figure}

Since all participants experienced the same single playback condition, we treated the midpoint of 4 on the 7-point scale as a neutral reference point and performed two-sided exact Wilcoxon signed-rank tests on the responses, testing whether responses to each item differed from this midpoint. This allowed us to examine whether responses were positively or negatively biased relative to a neutral reference point. The significance level was set at 0.05. Results significant at the more stringent 0.01 level are also noted below. To make these deviations from the midpoint of 4 easier to see, Figure~\ref{result} shows diverging stacked bar charts of the questionnaire responses.

\vspace{-2pt}
\subsubsection{Perceived Avatar-Mediated Drawing in Physical Space}
For Q1, ratings were significantly above the neutral midpoint ($W=0.000$, $p=0.008$). This result also met the more stringent 0.01 level. The interviews also showed that participants often saw the avatar as the drawer of the autograph. For example, one participant (P0) noted that \textit{``The avatar and marker movements were coordinated, so the avatar itself seemed to be doing the drawing''}, and another participant (P1) said that \textit{``The movements appeared synchronized, leading me to perceive the avatar as drawing''}. Overall, these interview comments matched the positive quantitative result for Q1.

Q2 responses tended to lie on the positive side of the midpoint, but they did not differ significantly from it ($W=0.000$, $p=0.063$). The interview comments were not all the same: some participants described the drawing as natural, whereas others pointed out visible timing mismatches between the avatar's hand and the marker. For example, one participant (P1) said, \textit{``During drawing, the avatar's hand and the marker appeared synchronized, so the marker stroke seemed to emerge naturally from the hand motion''}, whereas another participant (P2) noted \textit{``The marker moved ahead slightly at times, and I noticed the avatar's hand catching up''}. These comments added detail to the positive but non-significant result for Q2.

\vspace{-2pt}
\subsubsection{Naturalness of the Produced Autograph}
Ratings for Q3 fell significantly below the neutral midpoint ($W=0.000$, $p=0.008$). This result also met the more stringent 0.01 level. Participants' interview comments similarly suggested that the autograph drawn on the paper was not perceived as especially odd. For example, one participant (P3) said, \textit{``I was impressed by the high quality of the autograph the avatar produced''}, and another participant (P4) noted, \textit{``Some parts of the autograph were slightly shaky, but that did not bother me as something odd for an autograph''}. Together, these comments matched the quantitative result for Q3.

\vspace{-2pt}
\subsubsection{Perceived Avatar Presence During the Event}
Q4 also yielded ratings significantly above the neutral midpoint ($W=0.000$, $p=0.016$). Several participants described the avatar as being present in the physical setting throughout the event. For example, one participant (P0) said, \textit{``Although it was an avatar, I felt as if it was really there''}. This interview pattern matched the positive quantitative result for Q4.

No significant deviation from the midpoint was observed for Q5 ($W=1.500$, $p=0.094$). When asked whether the avatar's presence became stronger during drawing, participants gave different impressions. One participant (P6) said, \textit{``I felt as if the avatar was actually drawing in the real world, so I sensed a stronger presence during the marker strokes''}, while another (P4) noted, \textit{``Just having the avatar appear in real space already gave a strong sense of presence, so drawing did not feel like it became stronger''}. This difference in comments matched the non-significant result and the broad spread seen in Q5.

\vspace{-4pt}
\subsection{Discussion of Avatar-Mediated Drawing Event Plausibility}
In this subsection, we discuss whether the core Present Drawing experience in the current evaluation was experienced as an avatar-mediated drawing event.

\vspace{-2pt}
\subsubsection{Basic Plausibility of Avatar-Mediated Drawing Event}
Q1 and Q3 suggest that Present Drawing can support a basic form of an avatar-mediated drawing event, even in the playback condition used in this evaluation. Participants tended to feel that the autograph was drawn by the avatar and reported little oddness in the resulting trace. Together with the baseline sense of avatar presence observed in Q4, these findings suggest that the current system can make an avatar-mediated drawing event plausible by making the avatar appear to leave a physical trace in the real world.

\vspace{-2pt}
\subsubsection{Perceiving the Avatar as the Drawer Beyond Fine-Grained Synchrony}
Q1 showed a significant positive response, while Q2 responses were mixed and did not clearly exceed the neutral midpoint. Taken together, these findings suggest that, under this evaluation configuration, participants could still feel that the avatar was drawing even when perceived fine-grained synchrony was not uniformly strong. One possible interpretation is that participants experienced the avatar’s movement, the marker motion, and the emerging trace as one connected drawing action. At the same time, the interview comments related to Q2 indicate that noticeable timing mismatches can weaken this perception. Thus, while this sense remained plausible in the current configuration, reducing visible mismatches still appears important.

\vspace{-2pt}
\subsubsection{Avatar Presence in Drawing Event Content}
Q4 showed a significant positive response, but Q5 did not indicate that the avatar’s presence became stronger during drawing. Unlike prior work reporting increases in presence-related measures from avatar-linked physical effects \cite{leePhysicalVirtualTable2018,fabreKuroko2024}, this pattern may reflect methodological differences: prior studies compared conditions with and without avatar-linked physical influence, whereas our evaluation asked whether presence increased during drawing within a single condition. In this evaluation setup, the physical effect may have contributed more to perceiving the avatar as the drawer than to further increasing the avatar’s presence. Moreover, because Q4 and Q5 were included only as supplementary items rather than as a full assessment of an avatar's presence, this issue warrants further examination using validated measures such as Networked Minds \cite{bioccaDefiningMeasuring2002} or Basdogan’s Co-Presence Questionnaire (CPQ) \cite{basdoganExperimentalStudy2000}.

\vspace{-4pt}
\subsection{Discussion of Factors Supporting Perceiving the Avatar as the Drawer}
In this subsection, we discuss factors that the interview comments highlighted as important for seeing the avatar as the drawer in an avatar-mediated drawing event.

\vspace{-2pt}
\subsubsection{Lag and Drawing-Like Motion}
As discussed above, some participants commented on the lag between avatar motion and marker motion, suggesting that reducing visible lag may support seeing the avatar as the drawer. Participants also commented on the character of the marker motion itself. For example, one participant (P3) noted that \textit{``There were moments when it felt like a machine, rather than the avatar, was drawing because the stroke moved at an almost constant speed''}. This suggests that, in addition to spatial and temporal alignment, the motion profile of the marker may also affect this sense. Reproducing not only the stroke path but also drawing-like variations in speed and acceleration could help sustain the sense that the avatar is the drawer.

\vspace{-2pt}
\subsubsection{Mechanical Cues}
Interview comments also pointed to visible and audible signs of the mechanism as a possible influence on whether the avatar was seen as the drawer. Many participants commented that the marker could be perceived as being moved directly by the avatar because few visible mechanical cues were exposed during motion. At the same time, some participants noted that motor noise weakened this sense. For example, one participant (P3) said that \textit{``Because of the motor noise, there were moments when I felt that this stroke was being drawn by a machine''}. These interview responses suggest that concealing visible mechanical cues and reducing audible ones may help sustain the impression that the drawing is being produced by the avatar.

\vspace{-2pt}
\subsubsection{Real-Time Interaction}
While even a playback-based drawing event can support a basic sense of avatar-mediated drawing, interview comments also suggested that seeing the avatar as the drawer might change when participants could communicate with the avatar during the event and see the content of that interaction reflected immediately in the drawing. For example, one participant (P2) said, \textit{``If I could talk with the avatar and see my name or conversation content reflected on the spot, I think it would make the feeling that the avatar is the one drawing much stronger''}. This suggests that real-time interaction and on-the-spot personalization may help strengthen this sense.

\vspace{-4pt}
\section{LIMITATION \& FUTURE WORK}
\subsection{Initial Evaluation}
\vspace{0pt}
\subsubsection{Scope of the Current Evaluation}
The present evaluation was designed as a controlled first step to assess the visual plausibility of Present Drawing as an avatar-mediated drawing event. Accordingly, all participants observed the same playback condition: a single VTuber avatar performing the same short autograph-like drawing task, corresponding to one playback configuration with one avatar embodiment under human-operated control. Since the drawing-time synchronization between the avatar and marker is shared by the live and playback modes, the playback results remain informative about the visual plausibility of avatar-mediated drawing in both modes. However, although the visual impression may carry over, during drawing the avatar is driven to stay aligned with the marker to accommodate the CNC latency, so it lags behind the drawer's latest real-time pose. The evaluation does not yet show how this desynchronization affects the real-time communication between the drawer and participant in live conditions, and future evaluations should therefore examine how such live interaction shapes the experience.

Beyond this single playback configuration, it also remains to be shown whether similar impressions would be obtained with different avatar embodiments, in system-controlled configurations, or when the drawing task extends beyond short autograph-like content to more complex autograph-like or illustration-like drawings. The evaluation also used a single avatar and a small Japanese-speaking participant pool. Future evaluations should therefore examine a wider range of avatar designs and participant backgrounds, together with larger participant pools, to better understand how these factors shape the sense that the drawing is being produced by the avatar.

\vspace{-2pt}
\subsubsection{Mechanism Visibility and Display Method}
The current evaluation also examined Present Drawing only in the specific presentation setup used in our prototype: an HMD-based MR experience combined with a self-moving pen plotter whose visible mechanical structure was minimized. This setup was useful for assessing whether avatar-mediated drawing could be perceived under conditions designed to visually couple the avatar and the marker. However, we did not directly compare this setup against conditions in which the mechanism remained visible, so the effect of hiding the mechanical apparatus was not isolated in the present evaluation. 

In addition, the avatar was presented only through HMD-based viewing. Other presentation methods, such as projection-based display, may support different forms of shared viewing and accessibility, but their effect on how strongly the drawing feels as if it is being drawn by the avatar remains unexplored.

\vspace{-2pt}
\subsubsection{Field Evaluation}
The evaluation was conducted in a controlled laboratory setting with HMD-based observation and a fixed viewing position. It therefore does not yet show how Present Drawing would be received in more realistic social contexts. In particular, the appeal and reception of Present Drawing in actual fan events, exhibitions, or public installations remain unexplored. Future work should examine Present Drawing in field settings to understand how the experience changes in less controlled environments and in situations involving practical event use.

\vspace{-4pt}
\subsection{Prototype Implementation}
\vspace{0pt}
\subsubsection{Control of Self-Moving Pen Plotter}
In the evaluation, participants pointed out two related issues: small mismatches between avatar motion and marker motion, and a robot-like impression caused by nearly uniform motion. Both issues are likely related to the current grblHAL-based plotting setup. While grblHAL is suited to stably following a stroke path, it is not designed to reproduce handwriting together with its time-varying speed and acceleration. As a result, the marker motion in the current system can appear more uniform than human handwriting.

Another limitation is how avatar motion is synchronized with the pen plotter. During drawing, the avatar motion is driven in alignment with the CNC motion reported by grblHAL, rather than the original authored timing. However, when the marker is moving, the position reported by grblHAL can fall behind the actual marker position because it cannot keep up in time with the physical motion. As a result, a small mismatch can remain between avatar motion and the actual marker motion.

To address both issues, future systems may need time-aware closed-loop feedback control, such as proportional, integral, and derivative (PID) control, instead of the current grblHAL-based open-loop control. Combining such control with direct position sensing, for example through rotary encoders or computer-vision-based tracking, could help reduce motion mismatch and reproduce more drawing-like changes in speed and acceleration. Future work should also evaluate the technical performance of our self-moving pen plotter, which hides its actuation mechanism, including its ability to reproduce changes in drawing speed.

\vspace{-2pt}
\subsubsection{Drawing Media and Mount Constraints}
The current prototype used only one type of marker and one type of A4 Kent paper. Whether the same level of control and output quality can be maintained with other papers or other markers therefore remains unverified. In practice, changes in medium or tool may affect friction, stroke appearance, and controllability.

The current marker mount is also arranged so that a right-handed user can hold the marker in a natural posture. However, it does not easily support both right- and left-handed use with the same mechanism. One possible direction is to separate the visible marker held by the avatar from the physical marker that leaves the trace. For example, the avatar could hold a virtual marker object, while a small object with ink follows the virtual marker tip and leaves the trace on the surface. Such a design could make it easier to support different handedness and grip styles.

\vspace{-4pt}
\subsection{Future Directions}
The design space proposed in Section~3 is currently constrained by the form factor of the present prototype. Here, we outline several directions for extending the avatar-mediated drawing event beyond the current system.

One direction is to expand the scale of the event. Larger CNC systems could support drawing areas beyond A4 and enable more public or performative outputs. Another direction is to broaden the social structure of the event. For example, the one-drawer/one-observer setup could be extended toward collaborative drawing by combining Present Drawing with digital pens such as Neo Smartpen\footnote{Neo Smartpen: \url{https://neosmartpen.com}} that track stroke positions on specially printed paper. 

A further direction is to support multiple drawing avatars at once, enabling settings such as multi-character autograph sessions or collaborative compositions through coordinated actuators. Multi-robot pen-plotter systems such as iRobot Root\footnote{iRobot Root: \url{https://edu.root.samlabs.com/what-we-offer/root-robot}} provide one possible reference point for such extensions.

Present Drawing could also be extended in the range of physical tools it can present. The current prototype uses only one marker, but tool switching could allow avatars to change colors or marker types during an event. Existing automatic marker-changing mechanisms\footnote{Automatic Tool-Changing Pen Plotter Example: \url{https://howtomechatronics.com/projects/diy-pen-plotter-with-automatic-tool-changer-cnc-drawing-machine/}} suggest one possible direction. Supporting multiple tools could broaden the expressive range of Present Drawing.

\vspace{-4pt}
\section{CONCLUSION}
In this paper, we proposed Present Drawing, an approach for staging an avatar-mediated drawing event with physical traces in the real world. Our contributions were as follows:

\begin{enumerate}
  \setlength{\itemsep}{1pt}
  \setlength{\topsep}{2pt}
  \setlength{\parsep}{0pt}
  \item We presented a concept of an avatar-mediated drawing event with physical traces, and described a design space for expanding conventional drawing events through an avatar from an organizer's perspective.
  \item We developed a proof-of-concept prototype featuring a hidden magnetically actuated pen plotter that combines planar marker motion with magnetic pen-up/down control to generate physical strokes while maintaining avatar-marker visual coupling, and demonstrated five application examples based on the design space.
  \item We conducted an initial evaluation of one representative configuration, showed the basic plausibility of its experience as an avatar-mediated drawing event, and highlighted factors that support perceiving the avatar as the drawer.
\end{enumerate}

\noindent We also discussed limitations of the prototype and evaluation, including synchronization, media constraints, and field deployment, as well as future directions for extending the scale, social structure, and expressive range of the avatar-mediated drawing event.

\bibliographystyle{ACM-Reference-Format}
\bibliography{ref_py}

\end{document}